\documentstyle[pre,multicol,aps,epsf]{revtex}

\begin{document}
\draft

\title
{\Large \bf 
Symmetry-breaking on-off intermittency  under modulation:\\
Robustness of supersensitivity,  resonance  and information gain 
}

\author{ Bambi Hu$^{1,2}$  and  Changsong Zhou$^1$ }

\address 
{ $^1$ Department of Physics and Center for Nonlinear Studies, 
Hong Kong Baptist University, Hong Kong, China\\ 
$^2$ Department of Physics, University of Houston, Houston, Texas 77204}

\maketitle

\begin{abstract}

Nonlinear dynamical systems possessing  an invariant
subspace in the phase space and chaotic or stochastic motion  within the subspace often display
on-off intermittency close to the threshold of stability of  the subspace.   
In a class of symmetric systems, the intermittency is symmetry-breaking 
[Ying-Cheng Lai, Phys. Rev. E {\bf 53}.  R4267 (1996)]. 
We report interesting and practically important  universal  behavior 
of robustness of  supersensitivity, 
resonance and information gain 
in this class of systems when subjected to a weak modulation.  
While intermittent loss of synchronization may be harmful to application of high-quality 
synchronization of coupled chaotic systems, the features reported here may lead to interesting application 
of  on-off intermittency. 

\end{abstract}
\pacs{ PACS number(s): 05.40.-a, 05.45.-a}

\begin{multicols}{2}
\section{Introduction}
Nonlinear dynamical systems possessing an invariant subspace are of great interest, 
particularly when the system  motion within the subspace can be chaotic or stochastic.    
Examples include  chaotic  systems with symmetry~\cite{l,os}  or coupled chaotic systems~\cite{pc}. 
Bubbling~\cite{abs,hcp} and on-off intermittency~\cite{pst} are typical behaviors in  system close 
to a threshold of transverse stability of the subspace due to the fluctuative nature
of the local transverse Lyapunov exponent in different part of the subspace. In bubbling,
the subspace is stable on average, but local instability can result in large bursts away
from the subspace when it is perturbed. 
The sensitivity of this weak stability to parameter mismatch and noise has been 
studied by Pikovsky and Grassberger~\cite{pg}.  
 Intermittent loss of synchronization~\cite{hcp} in  
experiments with inevitable noise and parameter mismatch is undesirable in application of 
high-quality synchronization, such as to communication~\cite{co}.  
In on-off intermittency, the subspace is slightly
unstable on average, but local transverse attraction  may keep the dynamics very close 
to the subspace for a long period of time. Great attention has been paid to the  duration of 
this laminar period which exhibits universal power law distribution in a broad class of 
systems~\cite{pst}. Noise in the system  prevents its state from approaching the subspace 
close beyond the noise level, thus
has important effects on the laminar period distribution~\cite{phh} or the escape time~\cite{cl}. 
In noisy environment, bubbling and on-off intermittency are essentially the same  phenomenon. 

Sensitivity in nonlinear systems can be very useful for applications such as controlling global
dynamics of the system by local tiny perturbations~\cite{ogy}. It is interesting to ask  whether the 
sensitivity of on-off intermittency may lead to any potential application of the phenomenon. 
An observation is that  in a class of symmetric systems, on-off intermittency 
can be  symmetry-breaking~\cite{l}, namely, 
the  bursting behavior does not possess the system symmetry when the  system has two
symmetric but distinct attractors. This  Letter reports that a combination of the sensitivity 
of on-off intermittency
and the symmetry-breaking of the bursting can result in remarkable features  
in the systems subjected to a weak modulation signal in the noisy environment.   

\section{The model}
Let  $x(t)$ represent the distance of the
dynamics from the invariant subspace, and $x(t)>0$ and $x(t)<0$ 
denote the dynamics in the two symmetric components  respectively. 
In the symmetry-breaking systems, transitions  between  $x>0$ and $x<0$  can
only occur when $x(t)$ comes to the level of the weak noisy signal.
For system displaying appreciable laminar state, main features of the dynamics can be 
described by the general linear equation close to the subspace:  
\begin{equation}
\dot{x}(t)=[\lambda+\sigma_1\xi(t)]x(t)+ \sigma_2 e(t) + p s(t). \label{sys}
\end{equation}
Here  $\lambda$ is the transverse Lyapunov exponent of the subspace, and $\sigma_1\xi(t)$ 
with $\langle \xi(t)\rangle=0$  is the 
fluctuation of the local Lyapunov exponent due to the chaotic or stochastic  
 motion within the subspace. In general,
chaotic system has quickly decaying correlation, and in a large enough time scale $t$,
 $\xi(t)$ has an asymptotic Gaussian distribution.   
$e(t)$ is the additive white noise with level $\sigma_2\ll \sigma_1$ 
and $s(t)$ is a  weak modulation signal. 
The exact form of the signal is unimportant 
for the phenomena reported below, provided it varies on a time scale slower 
than the characteristic times of the systems. Here we consider $s(t)$ a random 
binary stream  ($\pm 1$ with probability 0.5) with a bit duration $T$. 
$p \ll \sigma_1$ is the 
amplitude of the signal, and $R=p/\sigma_2$ provides a natural measure of the signal-to-noise
ratio (SNR). 

In Ref.~\cite{cl}, Cenys and Lustfeld studied the statistical properties of
escape time of on-off intermittency subjected to noise by means of Fokker-Planck equation.
It has been shown that on-off intermittency is very sensitive to noise~\cite{phh,cl}.
We employ the same approach of  Fokker-Planck equation, 
and focus  on the property of amplification  of the  weak external signal $s(t)$ in the system.
Our results will demonstrate that the system is also very sensitive to the weak signal, and 
the amplification of the weak signal is robust to the additive noise. This sensitivity exhibits 
resonant behavior as the system parameters change.

The Fokker-Planck equation  for  Eq.~(\ref{sys}) is 
\begin{eqnarray}
\frac{\partial W}{\partial t}&=&-\frac{\partial }{\partial x}\left\{\left[(\lambda+\frac{
\sigma_1^2}{2})x+ps(t)\right]W\right\} \nonumber\\
&+&\frac{1}{2}
\frac{\partial^2}{\partial x^2}[(\sigma_1^2x^2+\sigma_2^2)W]. \label{FK}
\end{eqnarray}
In general, it is quite difficult to solve Eq.~(\ref{FK}) exactly. Let $T_0$ be the relaxation time
 of the system after  $s(t)$  switching from  $+1 \; (-1)$  to  $-1 \; (+1)$.  
If $T\gg T_0$, the probability distribution $W(t)$ 
can establish an approximate   static state during each bit of the input signal. 
Under the adiabatic approximation $T\gg T_0$, $\partial W/\partial t\approx 0$, and  
 the static solution  can be obtained  analytically as 
\begin{equation}
W(x)=C\left(x^2+\frac{\sigma_2^2}{\sigma_1^2}\right)^{(\alpha-1)/2}\exp\left[\frac{2ps(t)}{\sigma_1\sigma_2
}\arctan\frac{\sigma_1 x}{\sigma_2}\right],
\end{equation}
where $\alpha=2\lambda/\sigma_1^2$. For $|\sigma_1 x/A|\gg 1$ ($A=\max(p,\sigma_2)$),
\begin{equation} 
W(x)\approx C|x|^{\alpha-1}\exp\left[\frac{\pi ps(t)}{\sigma_1\sigma_2}{\hbox{sgn}}x\right]. 
\label{approx}  
\end{equation}

Now we see that the behavior of the system  can be divided into two regimes.      
One is $|x|\gg |\sigma_2 e(t) + p s(t)|$,  where the dynamics is governed  approximately by  
$\dot{x}(t)=[\lambda+\sigma_1\xi(t)]x(t)$. Let $z=\ln |x|$, then 
 $\dot{z}(t)=\lambda+\sigma_1\xi(t)$ 
which describes  a Brownian motion  with a constant drift $\lambda$ and diffusion constant $\sigma_1^2/2$. 
The nonlinearity of the system can be modeled by an effective reflecting boundaries 
of the Brownian motion at 
$\pm x_{b}$, which is of the order of $\sigma_1$. The probability density has 
a power  form $W(x)=|x|^{\alpha-1}$, but is asymmetric for $x>0$ and $x<0$ in the presence of $s(t)$. 
The  system can rarely perform transition between $x>0$ and $x<0$ in this regime  due to the 
symmetry-breaking property,
until it comes to  the other regime, where the noisy input $\sigma_2 e(t) + p s(t)$  dominates the dynamics 
and the system performs transition  between $x>0$ and $x<0$ frequently.  
The behavior of the system is  determined  by the competition between the diffusion 
and the drift  of the Brownian motion. If the drift time 
$t_b= \ln(\sigma_1/A)/|\lambda|$ is much  smaller than the diffusion time 
$t_d=2\ln^2(\sigma_1/A)/\sigma_1^2$, the drift is dominant over the diffusion, and the 
system will either come to a  metastable state induced by the noisy input for $\lambda<0$,  or
approach some state  away from the invariant subspace for $\lambda>0$. 
In both cases, the weak noisy input has no significant effects on the system behavior, 
i.e. the system is insensitive to the modulation. On the other hand,  
the diffusion is dominant for $t_b\gg t_d$, and  the system can have access to both  the 
level of the weak input  and the boundary of the nonlinearity, exhibiting 
typical on-off intermittency and  sensitivity to the weak modulation.    
Variation of the parameter $\lambda$ or $\sigma_1$ affects the competition between 
the drift and the diffusion, and  the system is  expected to display
optimal response to the weak modulation with resonant characterization.  

\section{Robustness of supersensitivity to the weak signal}  

Let us consider the diffusion dominant region close to the critical point 
of the stability of the subspace, e.g. $|\lambda|\ll 1$, $|\alpha \ln (x_{b}/A)|\ll 1$,
$\sigma_1\sim 1$, $p\sim 10^{-m} (m\gg 1)$. Employing the approximation in Eq.~(\ref{approx}), 
and the effective reflecting boundaries at $\pm x_b$, 
 the ensemble average  $\langle x(t) \rangle$ 
 is estimated as 
\begin{equation} 
\langle x(t) \rangle\approx s(t)\frac{ x_{b}}{\ln(x_{b}/A)}\tanh \frac{\pi R}{\sigma_1},
\label{ensav} 
\end{equation}
which in the noise-free limit $\sigma_2\to 0$, assumes the form 
\begin{equation}
\langle x(t) \rangle\approx  s(t)\frac{x_{b}}{\ln(x_{b})-\ln p}.
\end{equation}
A logarithmical dependence of $\langle x(t) \rangle$ on the input level $p$
means  an amplification of the weak signal $ps(t)$ with an factor 
$\langle x\rangle/p \sim 10^{m}/m \;(m\gg 1)$, 
i.e. the system exhibits {\sl supersensitivity} to extremely  weak modulation close to the critical point. 
This sensitivity was also reported in an overdamped Kramers oscillator with multiplicative noise free from
additive noise ($\sigma_2=0)$, which is a specific example in this class of systems~\cite{gp}.
In the absence of $s(t)$, the system produces symmetric bursting pattern with $\langle x(t) \rangle=0$; while the bursting 
pattern is reorganized to manifest the weak signal after it is fed into the system (see Fig. 1). 
The most interesting and practically important  property  is that the weak signal is manifested even 
buried in a relatively high level of noise, namely, the {\sl robustness of the supersensitivity}.  
This behavior originates from the {\sl symmetry-breaking of the on-off intermittency} in the system.

To demonstrate the above analysis, we employ the following  system in simulations~\cite{l}
\begin{eqnarray}
\ddot{y}&=&-\gamma \dot{y}+4y(1-y^2)+f_0\sin\omega t, \nonumber\\
\dot{x}&=&(a+by)\sin(x)-x +\sigma_2 e(t) +ps(t), \label{model}
\end{eqnarray}
where $y$ constitutes the forced Duffing chaotic oscillator.
With $\gamma=0.05, f_0=2.3$ and
$\omega=3.5$,  the Duffing system is chaotic and
$\sigma_1\approx 0.964b$. The nonlinearity of the variable $x$ is related to an experimental 
model of superconducting quantum interference device (SQUID)~\cite{zmb}.
However, we should stress that the specific form of the nonlinearity is of no importance for the phenomena.    
The transverse Lyapunov exponent of the invariant subspace $x=0$ is $\lambda=a-1$ due 
to the symmetry of the Duffing chaotic attractor.
Fig. 1 shows  typical behavior of the system and good  agreement between the analytical 
and  the simulation results for $\langle x(t) \rangle $ as a function of $R$.   
The agreement demonstrates that the general stochastic model in Eq.~(\ref{sys}) gives good account
for this type of system even though the motion in the subspace is deterministic chaos.    

\section{Resonant behavior}   

It is difficult to perform  generally a quantitative  analysis of the system response 
to the weak signal based on the linear dynamics in Eq.~(\ref{sys}) as parameters $\lambda$ or 
$\sigma_1$  changes, because the effective boundary $x_{b}$ changes with the 
nonlinearity  and  the linear dynamical model with an effective
reflecting boundary  is often not sufficient to capture the  dynamical property
if $\alpha$ is appreciately positive.  Moreover, as the parameters change, the  
relaxation time $T_0$ may become  comparable to the bit duration $T$, and the transient behavior 
plays an important role in the system response  and an adiabatic approximation
is not valid any more. 
To demonstrate the resonant properties, we rely on simulations with the system in Eq.~(\ref{model}),  
while the  Brownian motion model can provide  a qualitative understanding 
of the properties, thus showing that the properties are generic and  universal for a general 
class of the systems.

For a system with on-off intermittent output $x(t)$, 
the ensemble average $\langle x(t) \rangle$ and  
the correlation between $s(t)$ and $x(t)$ is relatively small even for the noise-free case $\sigma_2=0$, 
due to the power law fluctuation of $x(t)$.  
To better characterize the response of the system to the modulation, we transfer the output 
series $x(t)$ into 
a binary stream $X(t)$ by a threshold crossing process: suppose $x(t)$ becomes larger than a 
prescribed threshold $x_{th}$ 
at some moment, after that  $X(t)$ will keep at $X(t)=1$ until $x(t)$ crosses $-x_{th}$
at another moment; $X(t)$ will not switch back from  $X(t)=-1$ to $+1$ until $x(t)$ 
crosses $x_{th}$ again, and so on.  This binary presentation captures the most important feature
of the transition of the bursting pattern between the two symmetric attractors.  
$X(t)$  has a strong correlation with  $s(t)$ for  weak noise case $\sigma_2<p$ if the system is close 
to the critical point. The exact value of $x_{th}$ is not crucial for the properties  
described below.   In the following, we fix $p=10^{-7}$, $T=2000$ and $x_{th}=1$,   and 
take the cross-correlation 
function $C$ between $s(t)$ and $X(t)$ estimated  using  $10^4$ bits of a  random stream of 
 $s(t)$  to demonstrate the resonant behavior in the system.  

(1) {\sl With the change of $\lambda$}. For $\lambda$  rather below 
the critical point $\lambda=0$, the system has  a metastable state 
close to  
the level of the noisy input, and the diffusion is not strong enough to produce large bursts  
frequently enough, $C$ will be small. On the opposite, if $\lambda$ is rather above the 
critical point, the drift is also  dominant so that the system can seldom access to the level of the
weak modulation, and becomes insensitive to the  switching of $s(t)$ between $\pm 1$, resulting
in a small $C$ again. Close to the critical point, the system can access to the 
level of the weak signal  and produce  large bursts frequently due to strong enough diffusion.  The 
switching of the weak signal is ``sensed'' and manifested by asymmetrical bursting to $x>0$ and 
$x<0$, giving an optimal value of $C$.  
This precess is illustrated  by $C$ as a function of $\lambda$ in Fig. 2 for various $R$ values.

(2){\sl With the change of $\sigma_1$}.
For a small $\sigma_1$ where the drift is dominant over diffusion, the system is not sensitive to
the weak noisy input and $C$ assumes a small value. 
With the increase of $\sigma_1$, the diffusion becomes stronger,  
resulting  in a smaller relaxation time $T_0$ and more frequent large bursts, and the system becomes more 
sensitive to the weak input. 
In the noise-free case $\sigma_2=0$, the increased sensitivity  enables 
$X(t)$ to keep closer in phase to the weak signal $s(t)$ and $C$ approaches closer to $1.0$ if
the system maintains to work in the symmetry-breaking regime, and in general  a resonant behavior 
is not expected. 
The picture  becomes quite different if $\sigma_2\neq 0$. With  smaller $T_0$ and increased 
sensitivity, the system 
can keep up with and manifest more and more noise-induced  transitions in shorter time-scales, 
and the transition rate of  $X(t)$ between $\pm 1$  may  become much  higher than that of  $s(t)$, 
leading to a decreasing $C$.    
An optimal response is achieved when the diffusion is strong enough to become sensitive to the weak input
but not too strong  to manifest a lot of noise-induced transitions in short time-scales. Typical example 
of the system response as a function of $\sigma_1$  is shown in Fig. 3.

The above resonant behavior is similar to the conventional stochastic resonance  where  a dynamical
system displays increased sensitivity to a  subthreshold signal with an optimal level of additive  noise,
 see Ref.~\cite{ghjm} for an extensive review.  
Resonance occurs when a noise-controlled time-scale in the system matches that of the signal. In 
our system, the underlying  mechanism of  the resonant behavior is quite different. 
The sensitivity to an extremely weak signal is induced by the {\sl multiplicative} chaotic
or stochastic motion in the subspace. To achieve this sensitivity, it is required that the system  
is in the on-off intermittency regime so that it can become 
susceptible to the weak signal by coming close enough to the subspace, and  manifest and amplify it by
quick enough large bursts away from  the subspace with symmetry-breaking.  As system parameters $\lambda$ 
and $\sigma_1$ change,  a competition between these 
two factors leads to the resonant behavior. More interestingly, resonant behavior with respect to the
change of $\sigma_1$,  the level of the multiplicative chaos (noise),   necessarily occurs 
only in the presence of the additive noise due to the nature of this competition. 
These features are  generic and universal  in a general class 
of systems displaying on-off intermittency  with symmetry breaking.    
Multiplicative stochastic resonance has been   studied by 
Gammaitoni {\sl et al}~\cite{gmms} in  a multiplicatively driven bistable system 
with $\lambda=1$. 
In that case, the system is out of the  regime of on-off intermittency, and consequently 
cannot display the property of (super)sensitivity, and the resonance with respect to the change of $\lambda$ 
was not resported.

\section{Information gain}

Now  consider the system from the viewpoint of transmission and amplification
of a weak signal $ps(t)$ contaminated 
with channel noise $\sigma_2 e(t)$ through a system displaying on-off intermittency with symmetry-breaking. 
It is very interesting  and  practically important  that more information about the signal may  be obtained from the output 
$X(t)$ than from the noisy input $ps(t)+\sigma_2e(t)$ itself,  
besides the fact that the weak signal has been amplifyed to a level discernible  with a low resolution detector.   
To examine the information gain, we compare  $C$ with the correlation  between 
the signal $ps(t)$ and the total noisy input $ps(t)+\sigma_2e(t)$, i.e. $C_{in}=R/\sqrt{1+R^2}$. 
$C$, $C_{in}$ and their difference 
are shown in Fig.~4. 
$C$ comes to a saturated value for $R\geq 1$, where  the noise-induced  transitions between $x>0$ and $x<0$
in short time-scales are rarely manifested by large bursts.  
This value does not approach $1.0$ as $C_{in}$ due to  an average  time-delay between $X(t)$ 
and $s(t)$ induced by the relaxation time which is  longer for weaker signal. 
Clearly, appreciable  information gain is obtained by the system in a broad range of  
$R$, and an optimal gain is found at a certain $R$ value.  
If taking into account the effect of  the time-delay between $X(t)$ and $s(t)$, e.g. by defining 
$C_{\tau}$ as the maximum of the correlation between $X(t)$ and $s(t-\tau)$, 
the information gain region can be wider (Fig.~4).

To conclude, we demonstrate interesting  universal features of  robustness of supersensitivity, 
resonance and information gain
in a class of nonlinear system subjected to a weak modulation.  These systems present a new mechanism of 
resonant behavior compared to conventional  stochastic resonance.  While intermittent loss of synchronization may
be harmful for any applications employing high-quality synchronization~\cite{hcp},
the features found in this letter  are meaningful for  potential  applications of on-off intermittency.   
On-off intermittency  has been demonstrated in many experimental systems and we believe that the
behaviors  reported in this work can be tested in physical experiments.

This work is  supported in part by grants from the Hong Kong Research Grants Council (RGC) and the Hong Kong
Baptist University Faculty Research Grant (FRG).

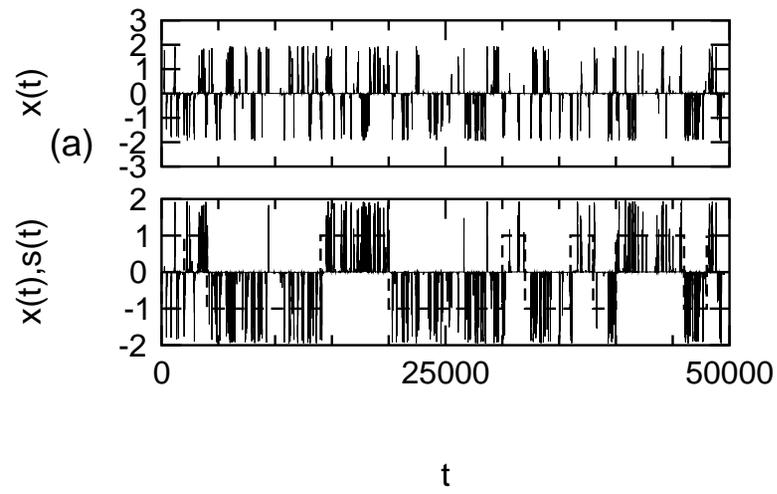
\begin{figure} 
\narrowtext
\caption{
(a) A typical time series of $x(t)$ in the noisy environment with 
$\sigma_2=10^{-6}$: upper panel, $R=0$, and  lower panel, $R=0.5$. 
The dotted line is $s(t)$. $\lambda=-0.02$, $T=2000$,     
(b) Numerical estimated $\langle x \rangle$ as a function of $R$ (stars) 
compared to the analytical estimation of Eq.~(\ref{ensav}) (line). 
$\lambda=0$, $b=1$ and $p=10^{-7}$. }
\end{figure} 

\begin{figure}
\narrowtext
\caption{
 Resonant behavior with respective to  $\lambda=a-1$; $b=1$.}
\end{figure}

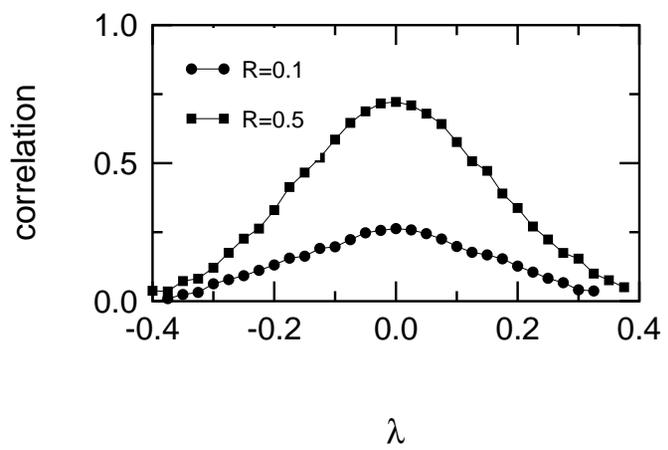
\begin{figure}
\narrowtext
\caption{
Resonant behavior with respective to $b$. 
A resonance occurs when $\sigma_2\neq 0$.}
\end{figure}

\begin{figure}
\narrowtext
\caption{
An illustration of information gain. $\lambda=0$, $b=1$. }
\end{figure}

\end{multicols}
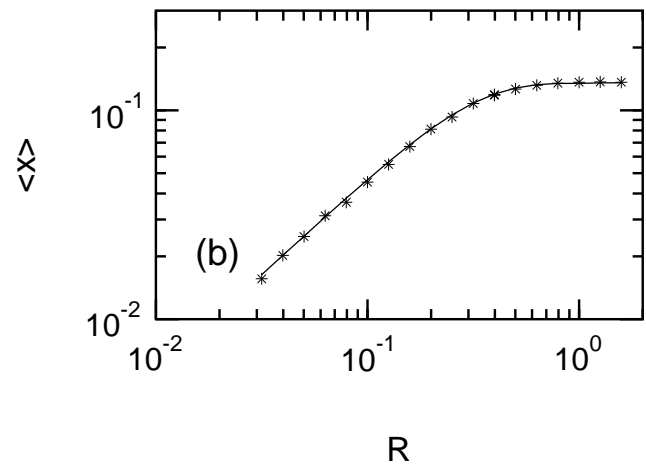
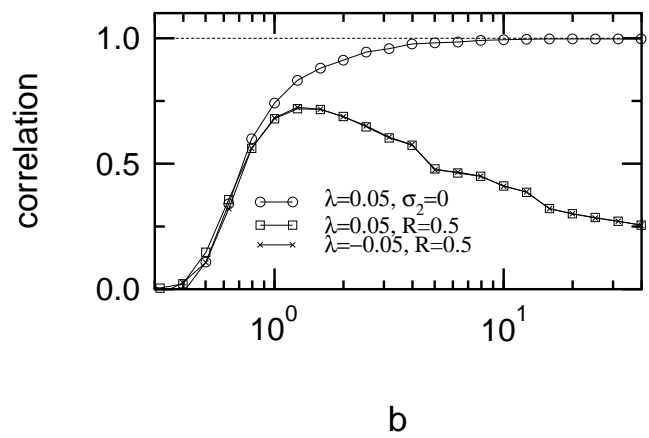
\end{document}